\documentclass[manuscript,sigconf]{acmart}

\newcommand{\avg}[1]{#1}

\newcommand{\minimalFeedback}[0]{\textit{Minimal feedback}}
\newcommand{\explanationFeedback}[0]{\textit{Explanation feedback}}

\newcommand{\XAI}[0]{\textit{AI recommendation and explanation}}
\newcommand{\update}[0]{\textit{Update}}
\newcommand{\explanationOnly}[0]{\textit{AI explanation only}}

\copyrightyear{2022}
\acmYear{2022}
\setcopyright{acmcopyright}
\acmConference[IUI '22]{27th International Conference on Intelligent User Interfaces}{March 22--25, 2022}{Helsinki, Finland} 
\acmBooktitle{27th International Conference on Intelligent User Interfaces (IUI '22), March 22--25, 2022, Helsinki, Finland}
\acmPrice{15.00}
\acmDOI{10.1145/3490099.3511138}
\acmISBN{978-1-4503-9144-3/22/03}

\begin{document}

\title{Do People Engage Cognitively with AI? Impact of AI Assistance on Incidental Learning}
\author{Krzysztof Z. Gajos}
\email{kgajos@eecs.harvard.edu}
\orcid{0000-0002-1897-9048}
\affiliation{%
  \institution{Harvard School of Engineering and Applied Sciences}
  \streetaddress{150 Western Ave.}
  \city{Allston}
  \state{MA}
  \country{USA}
  \postcode{02134}
}

\author{Lena Mamykina}
\email{om2196@cumc.columbia.edu}
\orcid{0000-0001-5203-274X}
\affiliation{%
  \institution{Columbia University}
  \streetaddress{622 West 168th St. PH-20}
  \city{New York}
  \state{NY}
  \country{USA}
  \postcode{10027}
}


\begin{abstract}

When people receive advice while making difficult decisions, they often make better decisions in the moment and also increase their knowledge in the process. However, such \em incidental learning \em can only occur when people cognitively engage with the information they receive and process this information thoughtfully. How do people process the information and advice they receive from AI, and do they engage with it deeply enough to enable learning? To answer these questions, we conducted three experiments in which individuals were asked to make nutritional decisions and received simulated AI recommendations and explanations. In the first experiment, we found that when people were presented with both a recommendation and an explanation before making their choice, they made better decisions than they did when they received no such help, but they did not learn. In the second experiment, participants first made their own choice, and only then saw a recommendation and an explanation from AI; this condition also resulted in improved decisions, but no learning. However, in our third experiment, participants were presented with just an AI explanation but no recommendation and had to arrive at their own decision. This condition led to both more accurate decisions and learning gains. We hypothesize that learning gains in this condition were due to deeper engagement with explanations needed to arrive at the decisions. This work provides some of the most direct evidence to date that it may not be sufficient to include explanations together with AI-generated recommendation to ensure that people engage carefully with the AI-provided information. This work also presents one technique that enables incidental learning and, by implication, can help people process AI recommendations and explanations more carefully. 
\end{abstract}




\maketitle

\section{Introduction}

In many areas of human enterprise, individuals increasingly rely on Artificial Intelligence (AI) to inform their decisions and choices. There is growing evidence that people supported by such systems can, on average, make better decisions compared to the decisions they would have made on their own~\cite{green2019principles,bussone2015role,lai2019human}. However, previous studies also showed human tendency to over-rely on AI-generated recommendations~\cite{bansal2021does,zhang2020effect,jacobs2021how,bucinca20:proxy}, which suggests that people may be processing information provided by AI superficially rather than engaging with it deeply and critically using their own knowledge and expertise. Given continuous concerns regarding the reliability and trustworthiness of AI, human critical engagement may be a necessary component of successful human-AI interaction, particularly in domains with a high cost of errors, such as health and medicine. There is already a growing body of work showing how the interactions with AI-powered decision support systems could be redesigned---using approaches ranging from tutorials~\cite{lai2020why} to in-the-moment cognitive interventions~\cite{bucinca2021trust,park2019slow}---so as to encourage deeper processing of the AI-generated information.

Researchers in learning sciences use the term ``cognitive engagement'' to describe learners' engagement with the learning process. When people are cognitively engaged with instructional process and materials, they are more likely to benefit from instruction and are more likely to acquire new skills and knowledge. We propose that cognitive engagement may be a useful construct in conceptualizing human engagement with AI and can help to distinguish between passive engagement, when individuals simply follow AI recommendations, and deeper forms of engagement, when they critically examine these recommendations and compare them with their own knowledge and judgement. An outcome of deeper cognitive engagement would be an ability to reject information that is inconsistent with individuals' own knowledge and beliefs, and to adjust their own knowledge to incorporate new information. This type of knowledge acquisition happens not only with formal instruction, but is also common in professional settings, when individuals interact with others in order to accomplish tasks, and use these interactions to increase their own knowledge ``about facts, domains, history, assumptions, strategies'' related to the task~\cite{berlin1992consultants}. This type of learning is commonly referred to as accidental learning~\cite{marsick01:informal,marsick2017rethinking}.

In this research, we examined the impact of different approaches to the design of human-AI interactions on incidental learning. Given previous research on human-AI interaction, we hypothesized that simply presenting a person with a decision suggestion and an explanation would provide an immediate benefit (i.e., help the person make a better decision) but would not lead to learning. However, we also hypothesized that alternative forms of the human-AI interactions---designed to elicit deeper processing of the AI-generated information---would both provide an immediate benefit and lead to incidental learning. We tested two such alternative designs. In the first one, which we refer to as the \update{} design, people first made an initial decision on their own before being shown the AI recommendation and explanation and having a chance to revise their decision. In the second one, the \explanationOnly{} design, participants were shown just an AI explanation but no explicit recommendation---they had to use the information from the explanation to arrive at the optimal decision themselves.

We conducted three experiments in which participants had to make a series of nutrition-related decisions (decide which of two meals shown was a greater source of a specified macronutrient). In each experiment, we compared one human-AI interaction to two non-AI baselines (in one baseline condition, participants received simple correctness feedback on their choices; in the second, they received both correctness feedback and an explanation). In all three experiments, the AI assistance provided significantly higher immediate benefit compared to the baseline conditions where such assistance was not present. As hypothesized, the results of the first experiment (n=251), showed that simply presenting people with an AI recommendation and explanation did not result in greater learning than in the baseline design where people received no assistance and no feedback. Contrary to our expectations, the results of the second experiment (n=268), showed that the \update{} design also did not result in learning. However, the results of the third experiment (n=221 and a replication with n=300) demonstrated that the \explanationOnly{} design did result in learning while also providing immediate benefit. 

We hypothesize that the observed difference in learning gain was due to the degree of cognitive engagement with AI-generated information. When individuals were provided with a solution to their task (in the form of a decision recommendation), they did not need to engage deeply with the explanations and could simply proceed with action. However, when they needed to arrive at their own decisions, they needed to engage with the provided explanations and synthesize the information to arrive at the conclusions. These results have implications for future AI-powered systems for supporting human decisions: contrary to common expectations, merely providing explanations for AI recommendations may not be enough to ensure that people critically evaluate those recommendations and arrive at final decisions that appropriately combine their own knowledge and information contributed by the AI. Instead, other forms of AI support that focus on presenting useful information rather than recommendations for solutions, may elicit deeper cognitive engagement and prompt individuals to more critically and thoughtfully examine assistance from AI. 

In this work, we make the following contributions:
\begin{itemize}
    \item The results of our first experiment show that people who were offered an AI-generated decision recommendation accompanied by an informative explanation performed better on the task at hand compared to when no AI support was offered, but did not learn from the AI-provided information. Given the strong link between cognitive engagement and learning, this is some of the most direct evidence to date that it may not be sufficient to include explanations together with AI-generated recommendation to ensure that people engage carefully with the AI-provided information.
    \item We demonstrated that an alternative design of the human-AI interaction, one in which the AI presents just an explanation leaving the person to arrive at the decision, provides both an immediate benefit in terms of decision quality and supports incidental learning.
\end{itemize}

\section{Related Work}
\label{sec:related}


Contrary to the initial expectations~\cite{kamar2012combining, kamar2016directions}, people supported by AI-powered decision support systems often make less accurate decisions on average than AI-powered systems on their own~\cite{bucinca20:proxy,bucinca2021trust,jacobs2021how,lai2019human,bussone2015role,bansal2021does,green2019principles,vaccaro2019effects}. This is surprising because if people combined their own knowledge with the information provided by the decision support systems, the resulting decisions should be more accurate than those made by either unaided people or AI-powered systems alone. A number of researchers investigated possible reasons for these surprising results. There is converging evidence that people overrely on the AI-generated recommendations~\cite{jacobs2021how,lai2020why,zhang2020effect,bucinca20:proxy} and that providing explanations for the AI recommendations might even exacerbate the problem~\cite{bansal2021does}. Recent work demonstrated that certain kinds of in-the-moment interventions---such as forcing people to wait before submitting a decision, having people state their initial decision before seeing the AI recommendation, or having the AI recommendation provided only on demand---can reduce (but not eliminate) this overreliance~\cite{bucinca2021trust}. It is possible that these interventions are effective because they encourage people to engage more deeply with the AI-provided information. We examine one of these interventions (the one where people state their initial decision before seeing the AI recommendation) in Experiment 2.


Research in cognitive psychology suggested that people process information on different levels. Deep processing occurs when individuals engage in more meaningful analysis of information and link it to existing knowledge structures~\cite{anderson1979elaborative}. In learning sciences, depth of processing is often associated with the degree of cognitive engagement, which is described as a ``psychological state in which students put in a lot of effort to truly understand a topic and in which students persist studying over a long period of time.''~\cite{rotgans2011cognitive}. 

While some authors discuss cognitive engagement as a personal trait of a student that does not depend on context~\cite{appleton2006measuring}, others suggest that cognitive engagement depends on the structure of each task ~\cite{rotgans2011cognitive,chi2014icap,greene2004predicting}. For example, searching for information on the Internet or engaging in discussion with other students engenders higher levels of cognitive engagement than passively listening to a lecture and results in higher learning gains. Rotgans and Schmidt attributes these differences in cognitive engagement to different degrees of autonomy afforded by different learning tasks~\cite{rotgans2011cognitive}. Chi et al. propose Interactive-Constructive-Active-Passive (ICAP) framework to describe a continuum of learning behaviors (from passive, to active, to constructive, to interactive) and argue that each subsequent level leads to an increase in cognitive engagement and learning~\cite{chi2009active,chi2014icap}. Further building upon the ICAP framework, Lam and Muldner showed that engaging in collaborative constructive activities has a positive impact on learning~\cite{lam2017manipulating}. 

In this study, we build upon these investigations in learning sciences and examine learning gains resulting from engagement with different types of AI-generated information. Previous research proposed a number of instruments for measuring cognitive engagement directly. However, many of the instruments developed thus far take a more longitudinal view of engagement and focus on completion of multiple activities involved in learning (such a rate of completion of homework, etc. as in~\cite{appleton2006measuring}). These instruments are ill-suited to capture variability in cognitive engagement within each individual task. Rotgans and Schmidt~\cite{rotgans2011cognitive} proposed an instrument more sensitive to the degree of cognitive engagement in real time; however, this instrument is still highly tailored to educational settings where learning is the primary goal, rather than brief tasks where learning is incidental. Consequently, in this study we did not use more direct measures of cognitive engagement, and instead we measured how much people learned.

Incidental learning typically occurs as a byproduct of other activities (e.g., problem solving, advice seeking) rather than as a result of explicit or formal educational activities~\cite{marsick01:informal}. However, like formal learning, incidental learning can only occur if people engage deeply with information. Incidental learning is common in professional settings~\cite{marsick01:informal,marsick2017rethinking}, and it can result from a wide range of professional activities and interactions, including colleagues assisting each other on decision-making tasks~\cite{berlin1992consultants}.

One recent project already demonstrated that people can learn from AI-generated recommendations and explanations provided that the explanations are carefully crafted to include relevant domain-specific information~\cite{das2020leveraging}. However, that project was conducted in the context of chess playing: participants received AI-generated recommendations and explanations on what moves to make. Because of the unique nature of the task (people who play chess typically are cognitively engaged in the game), it is unclear how broadly the results of that study generalize.

More extensive research on incidental learning has been done in an adjacent domain: navigational aids. A number of studies (e.g.,~\cite{munzer2012navigation,dey2018getting,bertel2017spatial}) demonstrated that the design of the navigation aid interface can influence both how well people can navigate in the moment and how much they learn about the route and the spatial configuration of the surrounding environment. While some researchers found trade-offs between how well different designs supported navigation and learning~\cite{munzer2012navigation}, Dey et al found an approach that supported both~\cite{dey2018getting}: They contrasted two designs that showed participants their location on the map. In one design, participants were also shown directional arrows telling them when and in what direction to turn, while the other design provided no such additional information. Both designs were equally effective at supporting navigation, but the latter design (the one without directional arrows) resulted in significantly greater learning than the one with the arrows. Generalizing beyond the domain of navigation support, these results suggest that individuals actively engage with and synthesize information when they need to arrive at a decision on their own; when the solution is presented to them as a suggestion, they process information more superficially, which inhibits learning. We build on this insight in Experiment 3.

\section{Experiment 1: AI Provides Recommendations and Explanations}
\label{sec:exp1}

The purpose of the first experiment was to evaluate whether incidental learning occurs when people receive decision support from a simple explainable AI system, one that offers a decision recommendation and an explanation. We hypothesized that such a design would improve people's immediate task performance compared to situations when no AI support was offered, but would not result in learning.

\subsection{Tasks and Conditions}
\label{sec:exp1tasks}

\begin{figure*}[p]
    \centering
    \includegraphics[width=.92\textwidth]{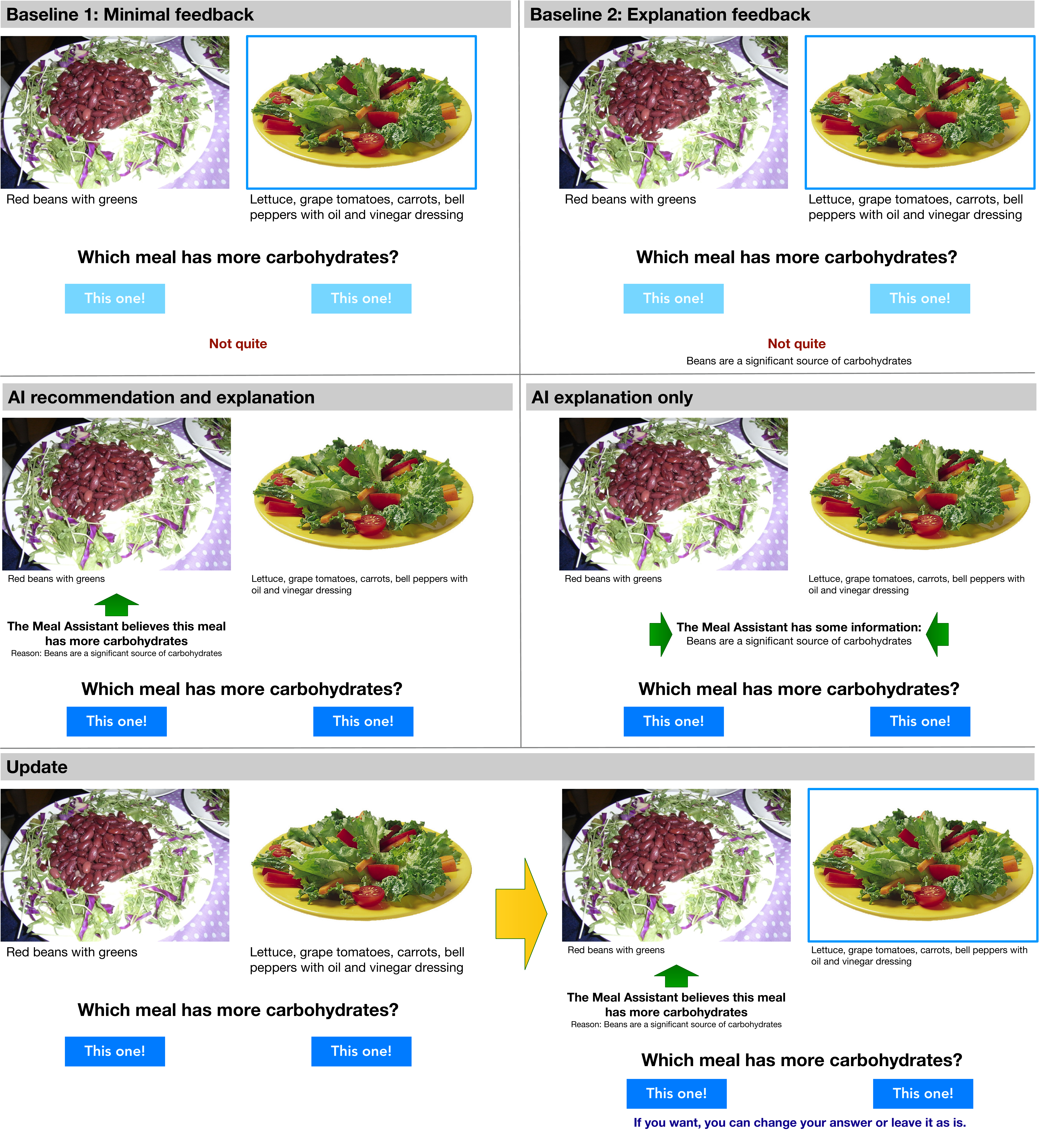}
    \caption{Experimental conditions included in the experiments. 
    Top-left: the \minimalFeedback{} design, a non-AI baseline condition where participants only receive correctness feedback (``Correct!'' or ``Not quite'') after they make their decision. 
    Top-right: the \explanationFeedback{} design, the second non-AI baseline condition, in which participants receive both correctness feedback and a brief explanation upon making their decision.
    Middle-left: \XAI{} design (Experiment 1)---participants see a recommendation and an explanation from a simulated AI systems, called the Meal Assistant, prior to making their own decision.
    Middle-right: \explanationOnly{} design (Experiment 3)---participants see only an explanation (but no recommendation) from a simulated AI system before making their own decision.
    Bottom: \update{} design (Experiment 2)---participants first make their own decision and then they are shown a recommendation and explanation from the Meal Assistant. They have the option to change their decision at this point.}
    \label{fig:conditions}
    \Description{Illustrations of all conditions show the same example with two meals. The meal on the left shows a picture and description ``Red beans with greens.'' The meal on the right is ``Lettuce, grape tomatoes, carrots, bell peppers with oil and vinegar dressing.'' Below the two meals is the question ``Which meal has more carbohydrates?'' and buttons for selecting either the left or the right meal. In the conditions where an explanation is shown, the following explanation is used: ``Beans are a significant source of carbohydrates.''}
\end{figure*}

To ensure that the results are informative, we used actual decision-making tasks rather than proxy tasks~\cite{bucinca20:proxy}. Specifically, we adopted the nutrition knowledge quiz (following the design of~\cite{burgermaster17:role}), in which participants were presented with images and descriptions of pairs of meals and were asked which meal contained more of one of the four macronutrients (carbohydrates, fat, protein, or fiber). Questions were designed such that each depended on one particular nutrition concept, such as that avocados are a significant source of fat or that soy beans contain more protein than most other beans. 

For each nutrition concept, we prepared three questions. They were used as a pre-test, the intervention, and post-test, respectively. 

The first question relating to a particular nutrition concept (the pre-test) was used to measure the participant's pre-existing knowledge of that concept. After responding to the pre-test question, they received only correctness feedback (``Correct'' or ``Not quite''). Prior work using this task design suggests that simple correctness feedback results in as little learning as no feedback at all~\cite{burgermaster17:role}, while our pilot data indicated that participants preferred receiving some feedback. Thus, providing simple correctness feedback minimized learning from the first exposure to the concept while improving participant experience.

The second time a participant encountered a particular concept (the intervention), they were presented with a design corresponding to one of the three experimental conditions (see the next section). By comparing participants' performance on the second exposure to a concept (and normalizing by their performance on the pre-test; see Section~\ref{sec:design_and_analysis_1}), we could estimate how much a particular design supported participants in performing the task at hand. 

The third question in each concept (the post-test) served as a means to assess how much participants learned about the concept from the intervention. After responding to the post-test, participants were presented both with correctness feedback and with an explanation --- this decision had no impact on the data collected and was, instead, intended to improve participants' experience.

The concepts were drawn at random from a larger pool, such that each study included two concepts for each of the four macronutrients (for a total of 8 concepts). Concepts were randomly assigned to conditions. The order in which questions were presented was randomized (thus, for each participant, different questions served as pre-test, intervention, and post-test). Consequently, the order of questions within each concept was randomized and the concepts were intermixed. There were a total of 24 questions (8 concepts~$\times$~3 questions per concept).

This experiment included three conditions: two baselines and a condition simulating a common approach to designing AI-powered decision-support tools:

\begin{itemize}
    \item \textbf{Baseline 1, \minimalFeedback.} In this condition, which is illustrated in Figure~\ref{fig:conditions}~(top left), participants received no AI assistance and only minimal correctness feedback (``Correct'' or ``Not quite'') after they submitted their answer. As mentioned before, in prior research this design was not significantly different from a design where no feedback at all was provided~\cite{burgermaster17:role}. Thus, we consider this to be the low baseline---it is unlikely that any other condition will result in less learning.
    \item \textbf{Baseline 2, \explanationFeedback.} In this condition, illustrated in Figure~\ref{fig:conditions}~(top right), participants again received no AI assistance, but they received both correctness feedback (``Correct'' or ``Not quite'') and a brief explanation (e.g., ``Avocados are a significant source of fat'') after they submitted their response. The same explanation was provided whether the participant answered correctly or not. In prior work, such feedback resulted in significantly greater learning than conditions where only correctness feedback was provided or where no feedback was provided~\cite{burgermaster17:role}. Thus, we consider this to be a high baseline: while an even more effective design might be possible, it represents a demonstrably effective solution. 
    \item \textbf{\XAI.} This condition (Figure~\ref{fig:conditions}~middle left) simulated the way AI-powered decision support systems are frequently implemented today: the AI recommendation accompanied by an explanation was presented at the very moment the person was presented with a decision task. To simulate real decision-making tasks where the ground truth is unknown, no feedback was provided to participants after they made their decision. Instead, they were told ``Your response has been recorded. (you will receive feedback at the end of the test)''. In the study instructions, the AI was introduced as the ``Meal Assistant, which is an experimental computer system that can analyze the nutritional content of meals.'' Additionally, they were told that ``The Meal Assistant is right most of the time but not always. You are welcome to consider its recommendations, but you should make whatever decision you think is best.'' This uncertainty is typical for contemporary AI-powered decision support, and can lead to different degrees of trust in AI-generated information. However, given that the focus of this study was on cognitive engagement, rather than trust, we designed our study such that the Meal Assistant recommendations were always correct.

\end{itemize}

The explanations presented in the \explanationFeedback{} and in the \XAI{} conditions were identical. They were also designed to take the form of \em contrastive \em explanations. Contrastive explanations show, for example, why a diagnosis should be disease X \emph{instead of} disease Y, where disease Y, used for contrast, is known as the \emph{foil}. When there are only two possible choices, the foil is implicit. Contrastive explanations include only information about what is relevant for choosing one option over the foil. There is broad consensus that contrastive explanations are among the most effective in human discourse~\cite{lipton1990contrastive,van1988pragmatic,miller2019explanation}. 
Also, the prior work that demonstrated that learning can occur in the context of the Nutrition Knowledge Test also used contrastive explanations (generated by an expert nutritionist)~\cite{burgermaster17:role}. Finally, recent work demonstrated that explanations that included explicit contrasts were more effective at causing people to select healthy meal alternatives than explanations that contained identical information about the meals but without the explicit comparison~\cite{musto2021:exploring}.

\subsection{Procedures}
Participants were recruited via two mechanisms: LabintheWild.org and Amazon Mechanical Turk (MTurk). LabintheWild is a platform for conducting online experiments with unpaid participants~\cite{reinecke15:labinthewild}. Instead of being paid, participants are incentivised by the promise that at the end of the study they will see their own results and compare themselves to other test takers. Both curiosity and opportunities for social comparison have been shown to increase engagement of online participants~\cite{law16:curiosity,huber17:effect} and multiple validation studies demonstrated that data collected on LabintheWild and other similar platforms are valid and lead to the same conclusions as data collected in traditional laboratory settings~\cite{germine12:web,reinecke15:labinthewild,huber20:conducting,li18vounteer,li2020controlling}. While some LabintheWild studies attracted tens of thousands of participants~\cite{reinecke14:quantifying,gajos20:computer}, experimenters have little control over the rate at which participants arrive. Thus, we supplemented recruitment with MTurk, which is also an effective choice collecting valid behavioral data~\cite{komarov13:crowdsourcing}. We paid MTurk participants \$1 (US) aiming for \$10/hour (the median time to complete the study was 6 minutes). 

Upon arriving at the experiment web site, all participants were presented with brief information about the study (including a promise to see their own results and the aggregate results of others at the end) followed by informed consent. Next, participants were asked if they had taken this study before and they were presented with a demographics form (where all questions were optional). Following~\cite{spiel2019:better}, we offered five options for gender: male, female, non-binary, prefer to self-describe (which enabled a free response field), and prefer not to answer. Then, they were presented with the instructions. Next, they completed the main nutrition knowledge test consisting of 24 questions. After the main test, they were asked if they had experienced any interruptions, technical difficulties, or if they cheated in any way. Finally, they were shown their own results, the average accuracy of other test takers, and the correct responses (and explanations) for all the questions in the test.

LabintheWild participants were also given an option to share the study via social media or to explore other studies hosted on LabintheWild. Meanwhile, MTurk participants were given given a verification code to enter back on the MTurk web site.

\begin{table*}[t]
\begin{tabular}{l|llll}
\toprule
\textbf{}       & \textbf{Experiment 1}                                                                                                   & \textbf{Experiment 2}                                                                                                 & \textbf{Experiment 3}  & \textbf{Experiment 3 replication}                                                                                                 \\
\toprule
\textbf{n}      & 251                                                                                                                     & 268                                                                                                                   & 221    & 270                                                                                                               \\
\midrule
\textbf{Source} & \begin{tabular}[c]{@{}l@{}}LabintheWild: 217\\ MTurk: 34\end{tabular}                                                   & \begin{tabular}[c]{@{}l@{}}LabintheWild: 184\\ MTurk: 84\end{tabular}                                                  & \begin{tabular}[c]{@{}l@{}}LabintheWild: 36\\ MTurk: 185\end{tabular} 
& \begin{tabular}[c]{@{}l@{}}LabintheWild: 300\\ MTurk: 0\end{tabular} \\
\midrule
\textbf{Age}    & 11--90, M=33, SD=14.4                                                                                                                       & 6--75, M=33, SD=14.6                                                                                                                     & 9--82, M=38, SD=12.6     & 11--81, M=28, SD=14.5                                                                                                                  \\
\midrule
\textbf{Gender} & \begin{tabular}[c]{@{}l@{}}female: 114\\ male: 112\\ non-binary: 4\\ self-described: 1\\ not responded: 20\end{tabular} & \begin{tabular}[c]{@{}l@{}}female: 139\\ male: 100\\ non-binary: 3\\ self-described: 1\\ not responded: 25\end{tabular} & \begin{tabular}[c]{@{}l@{}}female: 98\\ male: 114\\ non-binary: 3\\ self-described: 0\\ not responded: 6\end{tabular} & \begin{tabular}[c]{@{}l@{}}female: 137\\ male: 88\\ non-binary: 12\\ self-described: 1\\ not responded: 62\end{tabular}\\
\bottomrule
\end{tabular}
\caption{Participants}
\label{tab:participants}
\end{table*}

\subsection{Approvals}
This experiment (as well as the subsequent experiments reported in this manuscript) was reviewed and approved by the Internal Review Board at Harvard University, protocol number IRB15-2398.

\subsection{Design and Analysis}
\label{sec:design_and_analysis_1}
This was a within-subjects experiment with three conditions (\minimalFeedback{}, \explanationFeedback{}, and \XAI{}). 

We collected two dependent measures, separately for each condition:

\begin{itemize}
    \item \textbf{Immediate benefit.} We quantified the improvement at the intervention time compared to pre-test by computing normalized change $c$~\cite{marx2007normalized} (one of the measures commonly used to measure student improvement) between the average correct rate for intervention questions and the average correct rate for the pre-test questions:
    \begin{equation}
        c = \left\{
        \begin{array}{r|lr}
        \frac{\avg{\mbox{intervention}}\ -\ \avg{\mbox{pre}}}{1 - \avg{\mbox{pre}}} & \mbox{if}\ \avg{\mbox{intervention}} > \avg{\mbox{pre}} \\
        \frac{\avg{\mbox{intervention}}\ -\ \avg{\mbox{pre}}}{\avg{\mbox{pre}}} & \mbox{if}\ \avg{\mbox{intervention}} < \avg{\mbox{pre}} \\
        0 & \mbox{if}\ \avg{\mbox{intervention}} = \avg{\mbox{pre}}
        \end{array}
        \right.
    \end{equation}
    where ``intervention'' stands for the average correct response rate during intervention trials and ``pre'' denotes the average correct response rate during the pre-test trials.
    
    \item \textbf{Learning.} To quantify learning, we used an analogous approach to compute normalized change between the average correct rate at the post-test and the average correct rate at the pre-test.
    
\end{itemize}

Based on the results from the prior research that measured learning in the context of the Nutrition Knowledge Test~\cite{burgermaster17:role}, we powered our experiment to detect  differences corresponding to Cohen's $d \ge 0.2$. Given the within-subjects design of our experiment, this meant a minimum of 199 participants.

Our measures were not normally distributed. Therefore, we used a non-parametric test, Wilcoxon signed-rank test, to test for statistically significant differences in our data.

We computed effect sizes using a standard approach for Wilcoxon non-parametric tests~\cite{fritz2012effect,tomczak2014need}: $r=\frac{Z}{\sqrt{n}}$ where $Z$ is the test statistic produced by the Wilcoxon signed rank test and $n$ is the number of participants in the sample. The interpretation of this effect size follows Cohen's guidelines for $r$: .5 means a large effect, .3 a medium effect, and .1 is a small effect~\cite{cohen1988statistical}. 

\subsubsection{Treatment of Outliers}
We removed from the analyses all participants who indicated that they had taken the test before. Otherwise, we kept all the participants. However, we did analyze our data for extreme outliers to make sure that they did not improperly affect the results. Because our primary outcome measure was bounded (and extreme values were plausible), we used trial completion times as indicators of unusual behavior. Specifically, we flagged all trials that took more than 1 minute to complete and we repeated our analyses after removing all participants who had at least one outlier trial. All the conclusions in all the experiments still held after outliers were removed.

\subsection{Results}
\subsubsection{Participants}
251 people participated in this experiment. Their demographics are summarized in Table~\ref{tab:participants}.

\begin{figure*}
    \centering
    \includegraphics[width=\textwidth]{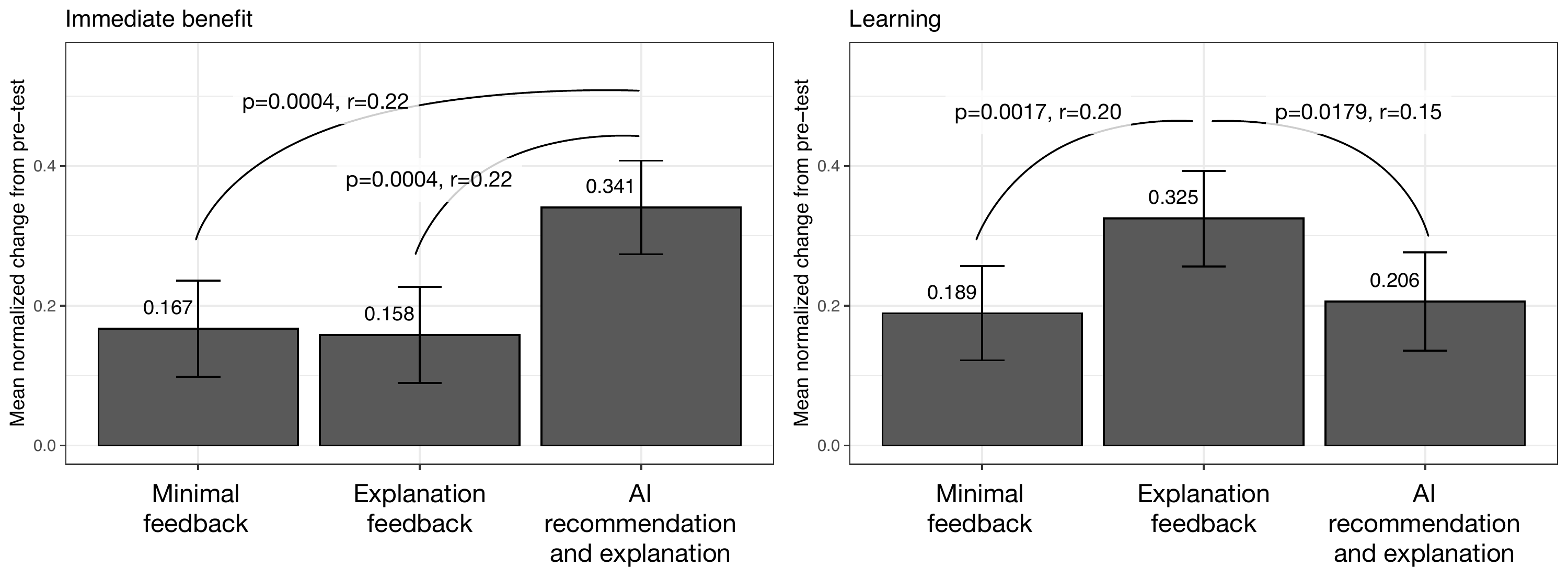}
    \caption{Experiment 1 results. Left: the immediate benefit per condition. Right: learning per condition. All results are reported as mean normalized changes from the pre-test. Error bars show 95\% confidence intervals.}
    \label{fig:exp1results}
    \Description{Two column graphs. Left: the immediate benefit per condition shows that people in the \XAI{} condition performed significantly better when AI assistance was offered compared to people in either \minimalFeedback{} or \explanationFeedback{} conditions, who received no AI assistance. Right: learning per condition shows that people in the \explanationFeedback{} condition learned significantly more than people in either \minimalFeedback{} or \XAI{} conditions.}
\end{figure*}

\subsubsection{Main Results}
The key results are visualized in Figure~\ref{fig:exp1results}.

As hypothesized, participants demonstrated a significantly higher immediate benefit of the intervention in the \XAI{} condition (normalized change between intervention and pre-test: M=0.341) than in either \explanationFeedback{} condition (M=0.158, Z=3.55, p=0.0004, r=0.22) or the \minimalFeedback{} condition (M=0.167, Z=3.56, p=0.0004, r=0.22). There was no significant difference in in terms of the immediate benefit between the \explanationFeedback{} and \minimalFeedback{} conditions (Z=0.10, n.s.).

The \explanationFeedback{} condition resulted in significantly larger learning gain (normalized change between post-test and pre-test: M=0.325) than the \XAI{} condition (M=0.206, Z=2.37, p=0.0179, r=0.15) or the \minimalFeedback{} condition (M=0.189, Z=3.14, p=0.0017, r=0.20). Consistent with our hypothesis, there was no significant difference in learning gain between the \XAI{} and the \minimalFeedback{} conditions (Z=0.68, n.s.). 
\section{Experiment 2: Participants Make Initial Decisions Before Seeing AI Recommendations And Explanations}

In this experiment, we tested if people experience the benefit of incidental learning if they first make their own decision, and only then are presented with the AI recommendation and explanation (and an option to revise their initial decision). We refer to this form of human-AI interaction as the \update{} design. In prior work the \update{} design resulted in more accurate decisions~\cite{green2019principles,fogliato2021impact} and reduced overreliance on the AI~\cite{bucinca2021trust} compared to the \XAI{} design. Buçinca et al~\cite{bucinca2021trust} hypothesized that the \update{} design induces people to engage cognitively with the AI-provided information. Thus, we hypothesized that the \update{} design would result both in improved task performance and learning.

\subsection{Tasks and Conditions}
We used the same task design as in Experiment 1. In this experiment, we had the following conditions:

\begin{itemize}
    \item \textbf{Baseline 1, \minimalFeedback.} Just like in Experiment 1. 
    \item \textbf{Baseline 2, \explanationFeedback.} Just like in Experiment 1. 
    \item \textbf{\update{}} This condition is illustrated in Figure~\ref{fig:conditions}~(bottom) and follows the general design used in prior research~\cite{bucinca2021trust,green2019principles,fogliato2021impact}. In this condition, participants first made a decision on their own and only then they were presented with the Meal Assistant recommendation and explanation. At this point they could, but did not have to, change their answer. The Meal Assistant recommendation and explanation was presented regardless of whether the participant's answer was correct. Participants did not know on which tasks they would receive the Meal Assistant recommendation after providing their initial answer. 
\end{itemize}

\subsection{Procedures, Design and Analysis}

The procedures, experiment design, measures, and analysis approach were the same as in Experiment 1 with one exception: immediately after the main test and before asking if they experienced any interruptions, we asked participants to fill out an abbreviated Need for Cognition questionnaire. Following~\cite{gajos17:influence}, we used a four-item subset of a common 18-item instrument~\cite{cacioppo84:efficient}. We elaborate the reasons for including this measure in Section~\ref{sec:audit}.

\subsection{Results}
\subsubsection{Participants}
268 people participated in this experiment. Their demographics are summarized in Table~\ref{tab:participants}.

\subsubsection{Main Results}
\begin{figure*}
    \centering
    \includegraphics[width=\textwidth]{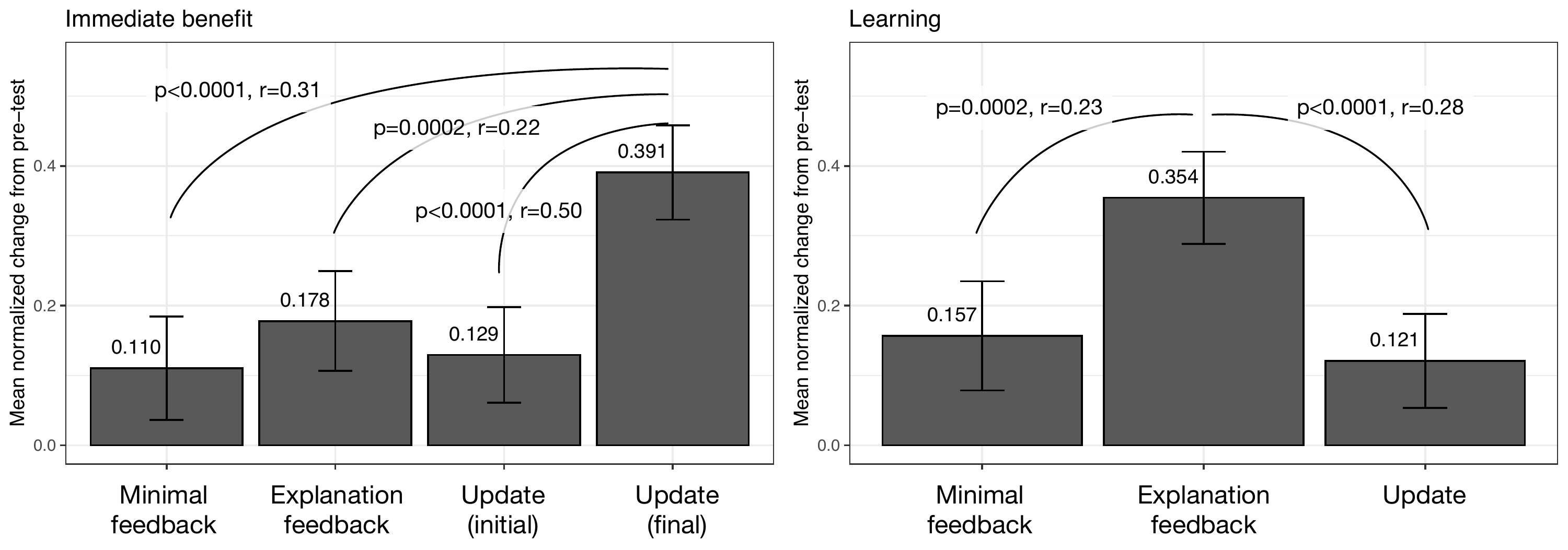}
    \caption{Experiment 2 results. Left: the immediate benefit per condition. Right: learning per condition. All results are reported as mean normalized changes from the pre-test (see Section~\ref{sec:design_and_analysis_1} for the definition of normalized change). Error bars show 95\% confidence intervals.}
    \label{fig:exp2results}
    \Description{Two column graphs. Left: the immediate benefit per condition. This graph shows two bars for the \update{} condition: one for participants' initial responses and one for their updated responses (after they saw the AI's recommendation). The graph shows that the final responses of people in the \update{} condition were significantly more accurate than people's initial responses in the \update{} condition or responses in either \minimalFeedback{} or \explanationFeedback{} conditions.  Right: learning per condition shows that people in the \explanationFeedback{} condition learned significantly more than people in either \minimalFeedback{} or \update{} conditions.}    
\end{figure*}

The key results are visualized in Figure~\ref{fig:exp2results}.

As hypothesized, participants experienced significantly higher immediate benefit in the \update{} condition (normalized change compared to pre-test M=0.391) compared to either the \minimalFeedback{} (M=0.110, Z=5.03, p<0.0001, r=0.31) or \explanationFeedback{} conditions (M=0.178, Z=3.67, p=0.0002, r=0.22). As expected, this improvement was due to participants changing their answers in response to the Meal Assistant's recommendations---their final answers were significantly more correct than their initial ones (M=0.129, Z=8.17, p<0.0001, r=0.50).  
There were no statistically significant differences in terms of the immediate benefit among the answers in the \minimalFeedback{} condition, the \explanationFeedback{} condition, or the initial answers in the \update{} condition.

Contrary to our expectations, participants did not learn more in the \update{} condition (M=0.121) than in the \minimalFeedback{} condition (M=0.157, Z=0.71, n.s., r=0.04). Participants learned significantly more in the \explanationFeedback{} condition (M=0.354) than in either \update{} (Z=4.59, p<0.0001, r=0.28) or \minimalFeedback{} (Z=3.76, p=0.0002, r=0.23) conditions.
\section{Experiment 3: AI Provides Explanations Only}

In this experiment, we evaluated the \explanationOnly{} design (Figure~\ref{fig:conditions}~middle right), in which people were presented just with the AI-generated explanation (e.g., ``Avocados are a significant source of fat'') but no explicit decision recommendation. As discussed in Section~\ref{sec:related}, this design was informed by the prior research on navigation aids, which suggested that when people are presented with the necessary information to make a decision but not with an explicit decision suggestion, they actively process the provided information resulting in good performance on the task at hand and in incidental learning~\cite{dey2018getting}. Consequently, we hypothesized that the \explanationOnly{} design would result both in improved task performance and learning.

\subsection{Tasks and Conditions}
We used the same task design as in Experiments 1 and 2. In this experiment, we had the following conditions:

\begin{itemize}
    \item \textbf{Baseline 1, \minimalFeedback.} Just like in Experiments 1 and 2. 
    \item \textbf{Baseline 2, \explanationFeedback.} Just like in Experiments 1 and 2. 
    \item \textbf{\explanationOnly{}.} This condition is illustrated in Figure~\ref{fig:conditions}~(middle right). Participants were presented with an explanation, but no recommendation as to which answer was correct. For example, participants were told that milk is a significant source of carbohydrates, but they had to process this information themselves to decide which answer to select. As in the \XAI{} and \update{} conditions, participants were not provided with any feedback on their answers on trial during which AI assistance was offered.
\end{itemize}

\subsection{Procedures, Design and Analysis}
All methods were the same as in Experiment 2 with one addition: after we conducted and analyzed the data from the experiment, we replicated it on an new independent sample. 

\subsection{Results}
\subsubsection{Participants}
221 people participated in the initial experiment and 270 participated in the replication. Their demographics are summarized in Table~\ref{tab:participants}.

\subsubsection{Main Results}
\begin{figure*}
    \centering
    \includegraphics[width=\textwidth]{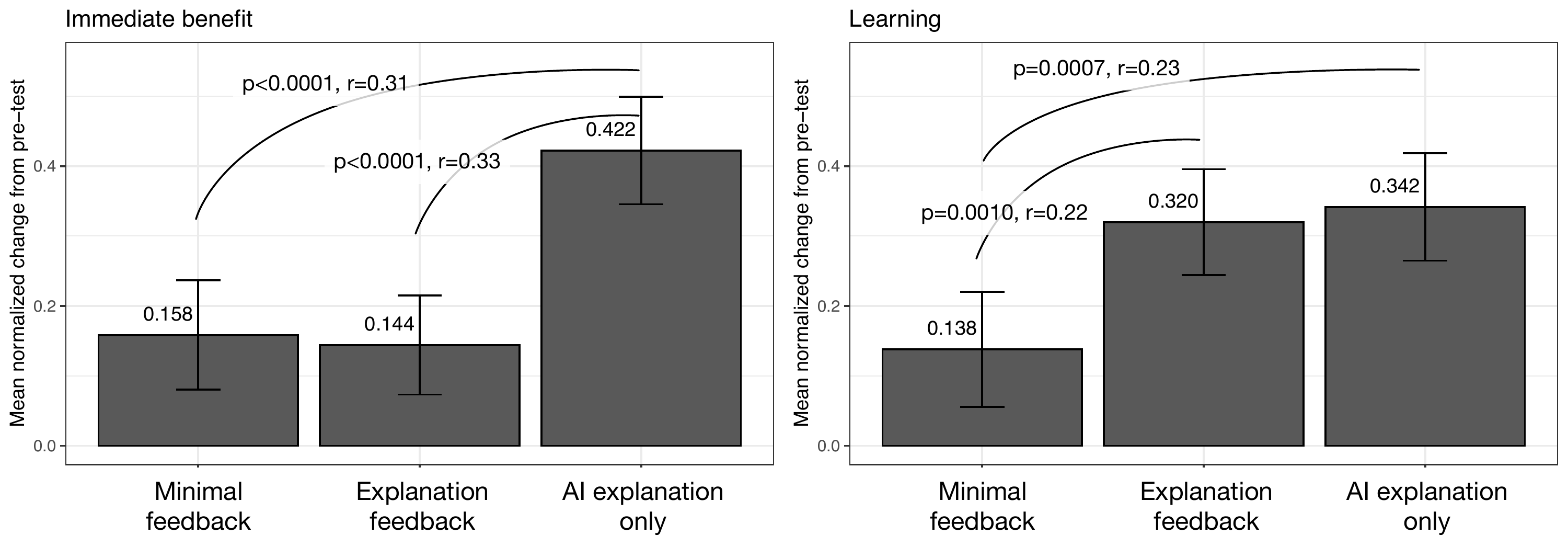}
    \caption{Experiment 3 results. Left: the immediate benefit per condition. Right: learning per condition. All results are reported as mean normalized changes from the pre-test (see Section~\ref{sec:design_and_analysis_1} for the definition of normalized change). Error bars show 95\% confidence intervals.}
    \label{fig:exp3results}
    \Description{Two column graphs. Left: the immediate benefit per condition shows that people in the \explanationOnly{} condition performed significantly better when AI assistance was offered compared to people in either \minimalFeedback{} or \explanationFeedback{} conditions, who received no AI assistance. Right: learning per condition shows that people in the \explanationFeedback{} and \explanationOnly{} conditions learned significantly more than people in the \minimalFeedback{} condition.}
\end{figure*}
\begin{figure*}
    \includegraphics[width=\textwidth]{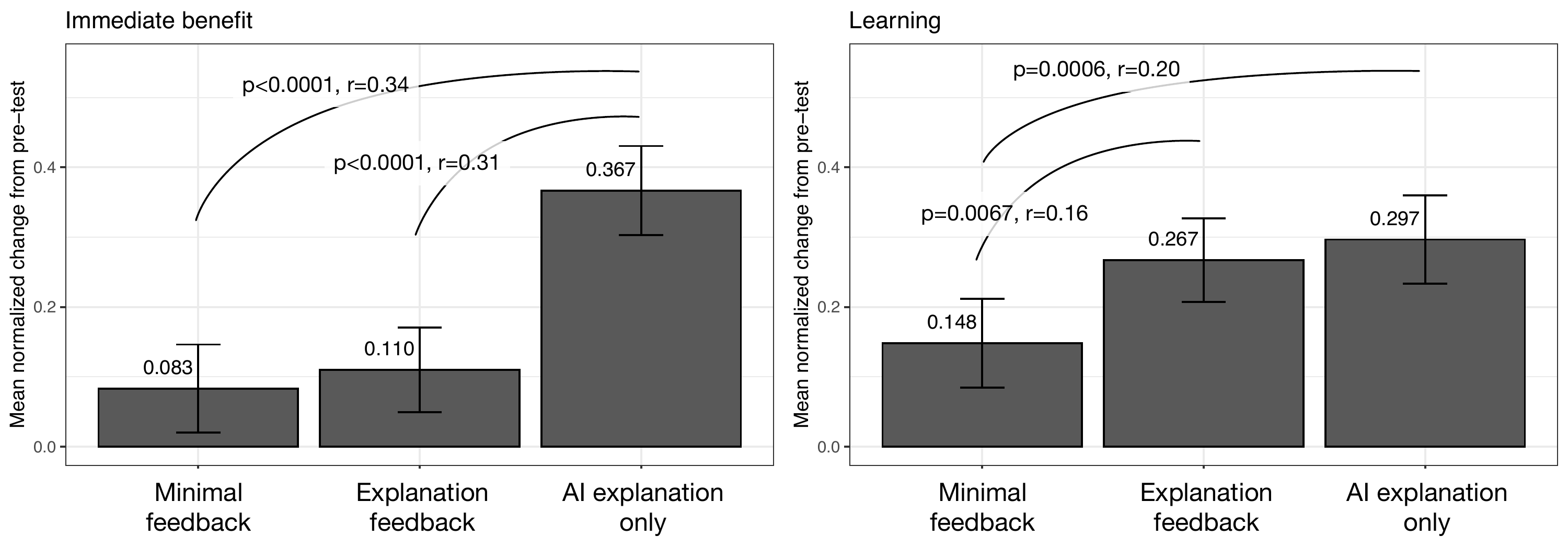}
    \caption{Experiment 3 replication results. Left: the immediate benefit per condition. Right: learning per condition. All results are reported as mean normalized changes from the pre-test. Error bars show 95\% confidence intervals.}
    \label{fig:exp3replication_results}
    \Description{Two column graphs. Left: the immediate benefit per condition shows that people in the \explanationOnly{} condition performed significantly better when AI assistance was offered compared to people in either \minimalFeedback{} or \explanationFeedback{} conditions, who received no AI assistance. Right: learning per condition shows that people in the \explanationFeedback{} and \explanationOnly{} conditions learned significantly more than people in the \minimalFeedback{} condition.}
\end{figure*}

The key results are visualized in Figure~\ref{fig:exp3results}.

As hypothesized, participants experienced greater immediate benefit in the \explanationOnly{} condition (M=0.422) than in either \minimalFeedback{} (M=0.158, Z=4.54, p<0.0001, r=0.31) or \explanationFeedback{} (M=0.144, Z=4.84, p<0.0001, r=0.33) conditions.

Also as expected, participants learned significantly more in the \explanationOnly{} condition (M=0.342) than in the \minimalFeedback{} condition (M=0.138, Z=3.39, p=0.0007, r=0.23). As before, participants also learned more in the \explanationFeedback{} condition (M=0.320) than in the \minimalFeedback{} condition (Z=3.28, p=0.0010, r=0.22). There was no statistically significant difference between \explanationFeedback{} and \explanationOnly{} conditions in terms of learning (Z=0.41, n.s., r=0.03).

The results of the replication (summarized in Figure~\ref{fig:exp3replication_results}) supported all of the conclusions from the initial experiment: all the significant differences remained significant with comparable effect sizes.

\subsection{Audit for Intervention Generated Inequalities}
\label{sec:audit}
\begin{figure*}
    \centering
    \includegraphics[width=.8\textwidth]{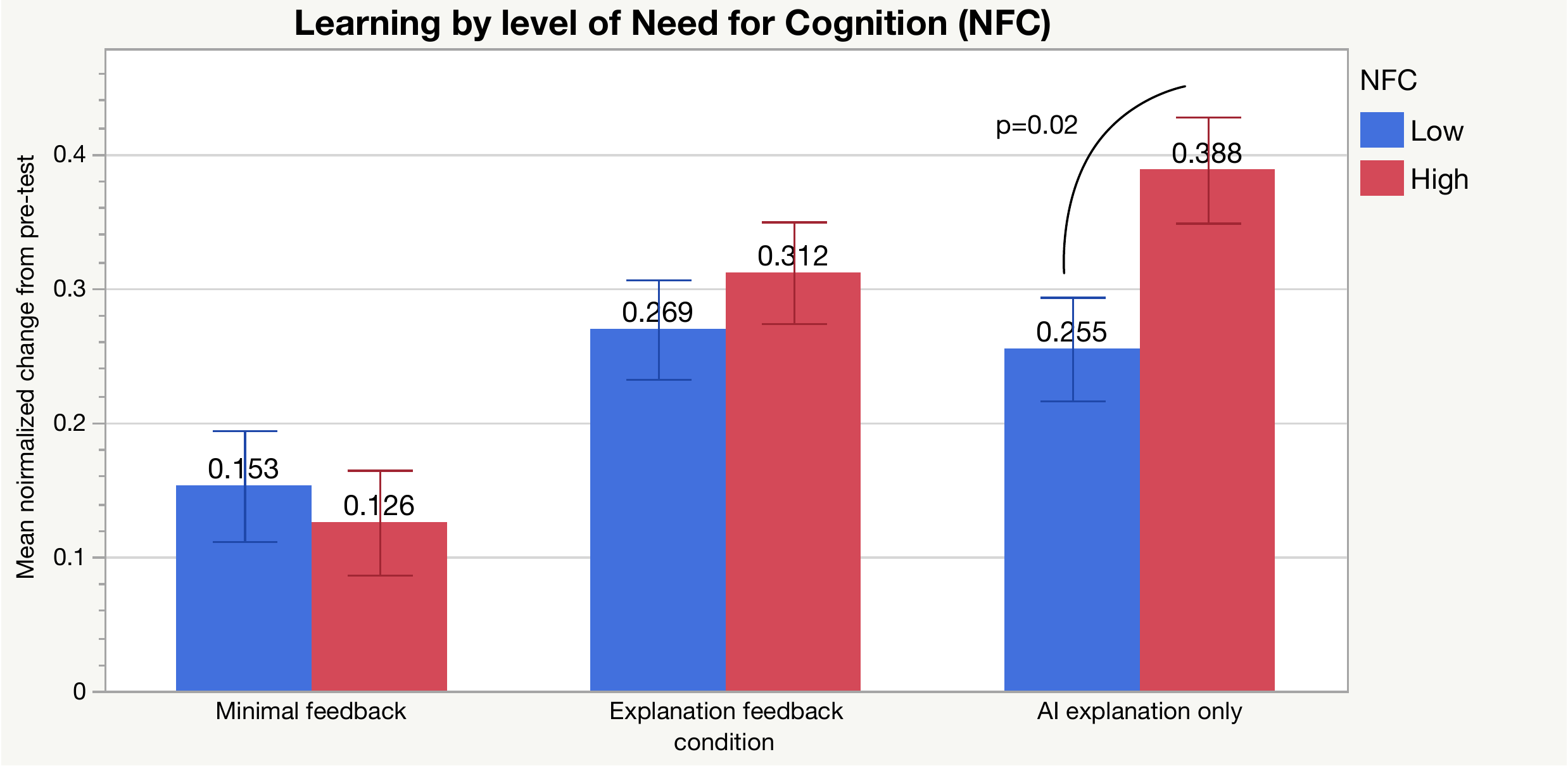}
    \caption{Results from Experiment 3 disaggregated by Need for Cognition. Error bars show 95\% confidence intervals.}
    \label{fig:exp3-nfc}
    \Description{Column graph showing three pairs of learning results: for \minimalFeedback{}, \explanationFeedback{}, and \explanationOnly{} conditions. For each condition, the results are shown separately for people with Low NFC and those with High NFC. There are no significant differences between the two NFC groups for the \minimalFeedback{} or \explanationFeedback{} conditions. For the \explanationOnly{}, however, the High NFC group learned significantly more ($p=0.02$) than the Low NFC group. }
    \end{figure*}
A design intervention, even if it is helpful on average, can be more useful to some groups than others resulting in \em intervention-generated inequalities\em~\cite{veinot2018good}. This is especially problematic if the intervention benefits an already privileged group more than others. An internal audit~\cite{raji2020closing}, which involves disagreggating the results by relevant demographic factors, can help uncover such problems. Given that our intervention targets the person's motivation to exert cognitive effort to engage with the AI-generated information, we built on similar prior work~\cite{bucinca2021trust} and disaggregated our results by Need for Cognition (NFC), a stable personality trait that reflects how much a person enjoys engaging in effortful cognitive activities~\cite{cacioppo82:need,petty86:elaboration}.

For this analysis, we used the data from both the original experiment and the replication. With NFC measured on a 1--5 scale, we divided our participants into two groups of roughly equal size: the Low NFC group (NFC $\leq 3.25$; n=229) and the High NFC group (NFC $>3.25$; n=207).

The results of our audit are illustrated in Figure~\ref{fig:exp3-nfc}. 
We did not observe statistically significant differences in learning between the two NFC groups in the \minimalFeedback{} condition (Wilcoxon rank sum test Z=0.668, n.s.) or in the \explanationFeedback{} condition (Z=0.677, n.s.). However, in the \explanationOnly{} condition the High NFC group learned significantly more (M=0.388) than the Low NFC group (M=0.255, Z=2.35, p=0.02). This result indicates that our intervention might have disparate effects for different individuals depending on their level of cognitive motivation.

\section{Discussion and Conclusion}

In this work, we examined how individuals engage with different types of AI-generated assistance and the impact of AI support on task performance and incidental learning. Specifically, we focused on the impact of decision recommendations and explanations, two increasingly popular components of AI-generated support with growing availability in both professional decision support tools~\cite{bates2021potential} and personal informatics systems~\cite{desai2019personal}. Previous research already demonstrated the positive impact of explanations provided as feedback on the task on incidental learning~\cite{burgermaster17:role}; as a result, we expected that explanations would be critical to learning. Furthermore, we were interested in examining interconnections between decision recommendations and explanations and their individual and combined impact on incidental learning.  

Our results showed that, as expected, all designs of the human-AI interactions we have tested provided significant immediate benefit, helping people make better decisions in those situations in which the AI assistance was offered. These results contribute to the growing body of evidence showing that people working with AI-powered decision support tools often (but not always~\cite{jacobs2021how,vaccaro2019effects}) make more accurate decisions than they would have on their own~\cite{green2019principles,bussone2015role,lai2019human}. 

As hypothesized, we observed the evidence of incidental learning in the \explanationOnly{} condition but not in the \XAI{} condition. As shown in Figure~\ref{fig:conditions}, in the \XAI{} condition participants were presented with both a decision recommendation and an explanation, whereas in the \explanationOnly{} condition they received just an explanation --- if they wanted to benefit from it, they had to process it carefully enough to infer which decision the explanation supported. While prior work has highlighted the critical role of explanations in promoting learning~\cite{burgermaster17:role,das2020leveraging}, our work additionally demonstrated the value of creating the conditions for learners to engage constructively (as defined in the ICAP framework~\cite{chi2009active,chi2014icap}) with the explanations.

Contrary to our hypothesis, we did not observe any evidence of learning in the \update{} design, as compared to the \explanationFeedback{} baseline. There are multiple possible explanations to this unexpected finding. First, when comparing this condition with the \explanationFeedback{} baseline, it is possible that differences in the attributed source of information between human experts (\explanationFeedback{} baseline) and AI (\update{} design) led participants to place different emphasis on otherwise identical information and engage with information provided by human experts deeper than with information provided by an AI. It is also possible that this lack of learning could be attributed to the framing of this information as either direct task feedback (\explanationFeedback{} baseline) or as additional information for contemplation (\update{} design) with people examining the task feedback more carefully than the optional additional information. Furthermore, it is possible that even though the participants were given a chance to update their decisions based on additional AI-generated information in the \update{} condition, they did not fully engage in the synthesis and simply accepted or rejected the recommended answer. This would suggest that this design did not fully reach the constructive level from the ICAP framework~\cite{chi2009active,chi2014icap}. Although this design has been previously shown to reduce overreliance on the AI recommendations~\cite{bucinca2021trust}, perhaps this effect was achieved by people reflecting more deeply on their own knowledge (and being more likely to follow their own judgement) rather than by carefully combining the AI-provided information with their own knowledge.

The finding that the \explanationOnly{} design appears to benefit people with high Need for Cognition (NFC) more than those with low NFC adds to the growing body of evidence suggesting that adding any form of ``intellligence'' to interactive systems generally benefits high NFC individuals the most~\cite{bucinca2021trust,carenini01:analysis,gajos17:influence,ghai2020explainable}. It is a potential source of concern because it suggests that the contemporary trends in interactive computing may be creating disparities that had not existed before.  

A key limitation of our work is that our simulated AI (the Meal Assistant) was always correct. This is a strength in the context of Experiments 1 and 2 as it shows that even in the idealized conditions the \XAI{} and \update{} designs did not support incidental learning. However, further work is needed to determine how sensitive the \explanationOnly{} design is to AI errors, and whether information on certainty of AI-recommendations can have an impact on engagement with explanations. A complementary future direction is to examine if the \explanationOnly{} design helps to reduce human overreliance on the AI. Another possible limitation is that our study included low-risk, inconsequential decisions and individuals had no special expertise in nutrition and nutritional judgment. Thus, these findings may not generalize to AI-assisted decision making by experts in domains with more consequential decisions; it is plausible that in those contexts individuals may exhibit deeper engagement with AI than was captured in our study. Furthermore, we only measured immediate learning but not long-term learning. While our results already indicate likely differences in the level of cognitive engagement across the three experiments, further work is needed to understand the long-term effectiveness of incidental learning from both expert and AI explanations. Lastly, we note that in the \explanationOnly{} condition the visual design emphasized the AI-generated explanations (with the green arrows pointing to the explanation) while in the other AI conditions the green arrows pointed to the correct answer. It is possible that by making the explanations more salient, the visual design of the \explanationOnly{} condition drew participants' attention to those explanations more than in the other AI conditions. However, we also note that explanations provided in the \explanationFeedback{} baseline robustly led to learning even though they were not visually salient. Thus, we are confident that the impact of the visual design in Experiment 3 was very small compared to the impact of the overall design of that condition. 

An important consideration in our work is that we used contrastive explanations~\cite{lipton1990contrastive,van1988pragmatic} and there is some evidence that the \explanationOnly{} design may be less effective at supporting the task at hand when non-contrastive explanations are used~\cite{lai2019human}. As previously explained in Section~\ref{sec:exp1tasks}, contrastive explanations answer the question why choose X instead of Y, where Y is known as the foil. Contrastive explanations are the dominant form of explanations in human-human discourse. They are much briefer than explanations that enumerate all evidence and thus make more efficient use of the cognitive resources of the person receiving the explanation. There is growing recognition that explainable AI systems should provide contrastive explanations~\cite{miller2019explanation,jacobs2021designing}, but this is not yet the norm. Recent work has produced several methods for efficiently computing contrastive explanations once the foil is known~\cite{alvarez2019weight,miller2018contrastive,van2018contrastive,krarup2019model,madumal2020explainable,dhurandhar2018explanations}. However, there remains the challenge of identifying a good foil. In some situations---when there are only two options to choose from or when the person reveals their initial choice---the foil is obvious. However, if we were to use the \explanationOnly{} design in a setting where many options are available (e.g., medical treatment selection~\cite{jacobs2021how}) new methods are needed to predict what options the user is likely to be considering so that the explanation provided addresses the right contrast.

Lastly, we note that there can be a trade off between encouraging deeper cognitive engagement and the efficiency or perceived usability of the different human-AI interaction designs. While it may be appropriate to design for deeper engagement in domains where the cost of errors is high and human expertise is readily available, such designs may incur higher cognitive burden and slower task completion. Future research need to explore trade-offs between different approaches to the design of AI-powered decision support.  

Our findings have important implications for the design of AI-powered decision support aids. First, they contribute compelling evidence in support of the emerging concern that people do not carefully process the AI-generated information when given a decision recommendation and an explanation~\cite{bansal2021does,bucinca2021trust,green2021flaws}. This, in turn, suggests an urgent need for research on fundamentally novel approaches to human-AI interaction. 

Second, our results suggest one such novel approach that places emphasis on synthesizing information necessary to arrive at decisions, but leaving the actual decisions up to human users. Our study provided strong evidence that this approach can promote deeper cognitive engagement with AI-generated information and can lead to higher quality decisions and learning. However, future research should examine trade-offs between cognitive burden associated with deeper engagement and task efficiency, whether similar approaches would hold in higher stakes domains, and the impact of uncertainty in the accuracy of AI-generated information on the degree of cognitive engagement. 

\begin{acks}
We thank Zana Buçinca, Maja Malaya, Barbara J. Grosz, Jean J. Huang, and the members of the Intelligent Interactive Systems group at Harvard for valuable feedback.
This work was funded in part by the National Science Foundation (grant IIS-2107391).
\end{acks}

\section*{Data Availability}
Data used in this research can be accessed at \url{https://doi.org/10.7910/DVN/JQL0JW}.

\bibliographystyle{ACM-Reference-Format}
\bibliography{kzg}

\end{document}